%% file: main.tex
\documentclass[twocolumn,showpacs,prl,amsfonts,amsmath,amssymb,floatfix]{revtex4} 

\usepackage[usenames,dvipsnames]{xcolor}
\usepackage{color}
\usepackage{hhline}
\usepackage{mathrsfs}
\usepackage{graphicx}
\usepackage{dcolumn}
\usepackage{bm}
\usepackage{multirow}
\usepackage{booktabs}
\usepackage{ulem}
\input{epsf}
\usepackage{amsmath}

\def\UH1{U^{{\rm lat}}}
\def\UDMFT1{U^{\rm{D\!M\!F\!T}}}
\def\UDMFTh1{U^{\rm{D\!M\!F\!T\mathchar`-high}}}
\def\UDMFTl1{U^{\rm{D\!M\!F\!T\mathchar`-low}}}
\def\BUDMFT1{{\mathbf U}^{\rm{D\!M\!F\!T}}}
\def\BUDMFTh1{{\mathbf U}^{\rm{D\!M\!F\!T\mathchar`-high}}}
\def\BUDMFTl1{{\mathbf U}^{\rm{D\!M\!F\!T\mathchar`-low}}}
\def\Zij1{Z_{\parallel \ \!\!\! i-j \ \!\!\! \parallel}}
\def\ij1{\parallel \ \!\!\! i-j \ \!\!\! \parallel}
\def\tij1{t_{ij}^{mn}}
\def\Uij1{U_{ij}^{mnop}}
\def\gs0ij{{{\mathbf G}^{0}_{ij,\sigma}}}
\def\Gs1ij{{{\mathbf G}_{ij,\sigma}}}
\def\g0ij{{{\mathbf G}^{0}_{ij}}}
\def\G1ij{{{\mathbf G}_{ij}}}
\def\Utij1{{{\tilde{U}}_{ij}^{mnop}}}
\def\Wh1{W^{\rm high}}
\def\Wl1{W^{\rm low}}
\def\Us1{U^{\rm scr}}
\def\bchi1{\mbox{\boldmath $\chi$}_{0}}
\def\chib2{\mbox{\boldmath $\chi$}}
\def\SrVO3{SrVO$_3$}
\def\mnop1{m\!n\!o\!p}
\def\ga{{\bf g}_1}
\def\gb{{\bf g}_2}
\def\gc{{\bf g}_3}
\def\gd{{\bf g}_4}
\def\gi{{\bf g}_i}
\begin{document}

\input{paper}

\clearpage
\input{supplement}

\end{document}

%% file: paper.tex
\title{Effective Onsite Interaction for Dynamical Mean-Field Theory}

\author{Yusuke Nomura$^{1}$}
\author{Merzuk Kaltak$^{2}$}
\author{Kazuma Nakamura$^{1,3}$}
\author{Ciro Taranto$^{4}$}
\author{Shiro Sakai$^{5}$}
\author{Alessandro Toschi$^{4}$}
\author{Ryotaro Arita$^{1,3,6}$}
\author{Karsten Held$^{4}$}
\author{Georg Kresse$^{2}$}
\author{Masatoshi Imada$^{1,3}$}
\affiliation{$^1$Department of Applied Physics, University of Tokyo, 7-3-1 Hongo, Bunkyo-ku, Tokyo, 113-8656, Japan} 
\affiliation{$^2$University of Vienna, Faculty of Physics and Center for Computational Materials Science, Sensengasse 8/12, A-1090 Vienna, Austria}
\affiliation{$^3$JST CREST, 7-3-1 Hongo, Bunkyo-ku, Tokyo, 113-8656, Japan}
\affiliation{$^4$Institute for Solid State Physics, Vienna University of Technology, A-1040 Vienna, Austria}
\affiliation{$^5$Centre de Physique Th\'eorique, \'Ecole Polytechnique, CNRS, 91128 Palaiseau Cedex, France}
\affiliation{$^6$JST-PRESTO, Kawaguchi, Saitama, 332-0012, Japan}
\date{\today}

\begin{abstract}
A scheme to incorporate non-local polarizations into the dynamical mean-field theory (DMFT) and a tailor-made way to determine the effective interaction for the DMFT are systematically investigated. 
Applying it to the 
two-dimensional Hubbard model, we find that non-local polarizations induce a non-trivial filling-dependent {\it anti-screening} effect for the effective interaction. 
The present scheme combined with density functional theory offers an {\it ab initio} way to derive effective onsite interactions for the impurity problem in DMFT.
We apply it to SrVO$_3$ and find that the anti-screening competes with the screening caused by the off-site interaction.
\end{abstract} 
\pacs{71.10.-w, 71.27.+a}
\maketitle 

{\it -Introduction}.
Understanding physical properties of strongly correlated electron systems is one of the most challenging subjects in condensed matter physics \cite{MACE, Rev-Kotliar}. For this purpose, it is essential to capture fermionic many-body effects necessitating a proper and accurate treatment of a large number of 
interacting fermions. The large number of electronic degrees of freedom in real materials 
are intractable, even with rapidly developing computational power. 
Hence, various ingenious ways of reducing 
the degrees of freedom have been developed. 
Aside from the reduction to mean-field effective one-particle Hamiltonians, 
 as in density functional theory (DFT), including dynamical fluctuations
for the reduced and tractable degrees of freedom 
is a route that has been explored extensively over the last decades. 

Approaches have been proposed \cite{MACE,Rev-Kotliar,Imai} to partially trace out the degrees of freedom far from the Fermi level, 
leaving an effective low-energy model for a small number of bands near the Fermi level. 
The resulting Hubbard-type lattice fermion models are much simpler than the original 
problem containing a huge number of bands. 
This reduction (downfolding) has been successfully incorporated in the constrained random phase approximation (cRPA) \cite{cRPA} by the use of maximally localized Wannier orbitals (MLWO) \cite{maxloc} as a basis set. It should be noted that, 
by tracing out certain electronic degrees of freedom, 
the effective interactions in the lattice fermion models (e.g. the Hubbard $U$, as exemplified by $U^{\rm cRPA}$ derived with the cRPA) are much reduced compared to the original bare Coulomb interactions \cite{cRPA-ex2,cRPA-ex3,cRPA-ex4,cRPA-ex5,cRPA-ex6,cRPA-ex7,cRPA-bedt,Ir,C60} because of the screening by polarizations of the eliminated degrees of freedom.

Although several efficient ways to solve the lattice fermion models
have been proposed \cite{MACE}, it is still too difficult to treat 
realistic situations so that a further reduction 
is highly desired.     
The widely used dynamical 
mean-field theory (DMFT) 
\cite{MV,KG} indeed offers a 
practical way of describing local correlation effects along this line \cite{Rev-Kotliar}, where 
the lattice fermion models are mapped onto quantum impurity models.

Although $U^{\rm cRPA}$ is widely used as input for DMFT calculations,
the conventional cRPA treatment totally excludes non-local screening processes
within the target band. These are also not contained in the DMFT, which only
accounts for the local screening processes. Hence, in the present work we argue
that a better starting point is the inclusion of non-local screening processes
of the target band within the RPA yielding an effective onsite
interaction $\UDMFT1$. 
Albeit tailor-made interaction parameters for the impurity problem were
employed in Ref.~\onlinecite{Kutepov}, a systematic investigation has been missing so far.
 
In this Letter, we examine a scheme for the systematic
determination of the effective onsite interaction
$\UDMFT1$ for DMFT calculations. This scheme is applied to both the 
two-dimensional (2D) single-band Hubbard model and 
to SrVO$_3$ 
by
using an {\it ab initio} description.
The application to the Hubbard model unexpectedly reveals the inequality $\UDMFT1 > U$ and a non-trivial filling dependence of $\UDMFT1$ with a peak around the van Hove singularity. A filling-dependent $\UDMFT1$ is also 
observed in the {\it ab initio} results for \SrVO3.
These are ascribed to an anti-screening effect induced by non-local polarizations, namely, a test-charge electron induces an off-site hole 
or electron and they again induce an onsite electron. 
This nonlocal effect increases $\UDMFT1$. 
The present elucidation contributes not only to the specific determination of the DMFT-interaction parameters, but also to gain insight into the nature of the reduced and simplified fermionic models in general. 

{\it -Equations to derive $\UDMFT1$}. 
Here, we 
derive the basic equations to evaluate $\UDMFT1$ from first principles calculations~\cite{Kutepov}. 
In the RPA, the screened Coulomb interaction $W$ can be written as $(1-v\chi_0)^{-1}v$
with the independent-particle polarization $\chi_0$ and the bare Coulomb interaction $v$.
The polarization $\chi_0$ is divided into $\chi_0^t$ and $\chi_0^r$, where $\chi_0^t$ is a polarization formed in the {\it target} subspace 
and $\chi_0^r$ is the rest. Note that this decomposition is not necessarily restricted to bands (cRPA); it is also applicable to the real space using localized basis sets. For example, the ``dimensional downfolding" has been formulated to derive effective models in reduced dimensions such as 2D or 1D models by excluding polarizations within the target layer/chain~\cite{cRPA-ex4}. With this decomposition and within the RPA, the fully screened $W$ can be obtained in a two-step procedure as~\cite{cRPA}
\begin{eqnarray}
\bar{W}=(1-v\chi_0^r)^{-1}v
\label{Wbar}
\end{eqnarray}
and 
\begin{eqnarray}
W=(1-\bar{W}\chi_0^t)^{-1}\bar{W},
\label{W}
\end{eqnarray}
where $\bar{W}$ describes a screened Coulomb interaction excluding 
a specified subset of excitations $\chi_0^t$. These excitations are
taken into account when the effective model with the interaction $\bar{W}$ is
solved. 
Alternatively, $\bar{W}$ is obtained from the fully screened $W$, 
by rewriting Eq.~(\ref{W})~\cite{Kutepov} as
\begin{eqnarray}
{\bar{W}}= W\bigl( 1+\chi_0^t W \bigr)^{-1}.
\label{Wbar2}
\end{eqnarray} 
In the present scheme, $\bar{W}$ corresponds to $\UDMFT1$ and  
$\chi_0^t$ is a one-center or local target polarization formed at the impurity site.

In practice, the static independent-particle polarization formed in 
the target bands ({\it tb}) is 
calculated using
\begin{eqnarray}
\label{DMFTeq11}
\chi_{0}^{tb} \! ({\mathbf r}, \!{\mathbf r}'\!) \!\!=\!\! 2 \! 
{\sum_{\alpha\beta}^{\in tb} } {\sum_{{\mathbf q}{\mathbf k}}}
\frac{f_{\beta\!{\mathbf k}\!+\!{\mathbf q}} \!\!-\!\! f_{\alpha\!{\mathbf k}} } { \epsilon_{\beta\!{\mathbf k}\!+\!{\mathbf q}} \!\!-\!\! \epsilon_{\alpha\!{\mathbf k}}}
\psi\!_{\alpha\!{\mathbf k}}^{\ast}\!(\!{\mathbf r}\!) \!
\psi\!_{\beta\!{\mathbf k}\!+\!{\mathbf q}}\!(\!{\mathbf r}\!)\!
\psi\!_{\beta\!{\mathbf k}\!+\!{\mathbf q}}^{\ast}\!(\!{\mathbf r}'\!) \!
\psi\!_{\alpha\!{\mathbf k}}\!(\!{\mathbf r}'\!), 
\end{eqnarray}
where \{$\psi_{\alpha{\mathbf k}},\epsilon_{\alpha{\mathbf k}}$\} are 
one-body wavefunctions and their energies with the wave vector ${\mathbf k}$ and the band index $\alpha$. The factor of 2 comes from the spin sum. The band summation is performed only over the target bands in the effective model.
Since the Bloch wavefunctions are related to
the Wannier functions via the unitary transform as 
\begin{eqnarray}
\label{DMFTeq12}
\psi_{\alpha\!{\bf k}}({\bf r})\!=\!\frac{1}{\sqrt{N}}\sum_{mi{\bf R}}
e^{i{{\bf k}} \cdot {\bf R}}
U^{\dagger ({\bf k})}_{mi,\alpha} \phi_{mi{\bf R}}({\bf r}), 
\end{eqnarray}
the polarization can be  
recast as 
\begin{eqnarray}
\label{DMFTeq13}
\chi_{0}^{tb}({\mathbf r},\!{\mathbf r}^{\prime}) &=& \frac{2}{N^{2}} \!
\sum_{mnop}  \sum_{ijkl}  \sum_{{\mathbf R}\!_{1}\!\mathchar`- {\bf R}\!_{4}}
\!\Biggl[  
\sum_{\alpha\beta}^{\in tb} \sum_{{\mathbf q} {\mathbf k}}
\frac{f_{\beta\!{\mathbf k}\!+\!{\mathbf q}}\!\!-\!f_{\alpha\!{\mathbf k}} }
{ \epsilon_{\beta\!{\mathbf k}\!+\!{\mathbf q}}\!\! -\! \epsilon_{\alpha\!{\mathbf k}} } 
e^{-i{\mathbf k}\cdot ({\mathbf R}\!_{1} \!-\! {\mathbf R}\!_{4} \!) }  \nonumber \\ 
&&\times  e^{i({\bf k}\!+\!{\mathbf q})\cdot ({\mathbf R}\!_{2}\! - \!{\mathbf R}\!_{3}\!)}
\!\Bigl(U^{\dagger ({\bf k})}_{mi,\alpha}\Bigr)^{\ast}
\!U^{\dagger ({\bf k}\!+\!{\bf q})}_{nj,\beta}
\!\Bigl(U^{\dagger ({\bf k}\!+\!{\bf q})}_{ok,\beta}\Bigr)^{\ast}
\!U^{\dagger ({\bf k})}_{pl,\alpha} \Biggr] \nonumber \\ 
&&\times \phi^{\ast}_{mi{\bf R}\!_{1}}\!({\bf r}) \phi_{nj{\bf R}\!_{2}}\! ({\bf r})
\phi^{\ast}_{ok{\bf R}\!_{3} }\!({\bf r}^{\prime}) \phi_{pl{\bf R}\!_{4}}\!({\bf r}^{\prime}), 
\end{eqnarray}
where $m$-$p$, $i$-$l$, ${\mathbf R}_{1}$-${\mathbf R}_{4}$ are the
orbital, primitive site, superlattice site indices respectively and $N$
indicates the total number of superlattice sites. With this expression, we specify the target-band polarization formed at the impurity site (the 0th site in ${\mathbf R}$=${\mathbf 0}$) as
\begin{eqnarray}
\label{DMFTeq14}
\chi_{0}^{{\rm im\!p}}\!({\bf r},\!{\bf r}^{\prime}\!) \!=\!\!\! \sum_{mnop}\!\! C_{m\!n\!o\!p} 
\phi^{\ast}_{m0{\mathbf 0}}(\!{\bf r}\!) \phi_{n0{\mathbf 0}} (\!{\bf r}\!) 
\phi^{\ast}_{o0{\mathbf 0}}(\!{\bf r}^{\prime}\!) \phi_{p0{\mathbf 0}}(\!{\bf r}^{\prime}\!) 
\end{eqnarray}
with 
\begin{eqnarray}
\label{DMFTeq15}
C_{m\!n\!o\!p} \!\!=\!\! \frac{2}{N^2}\!
\!\sum_{\alpha\beta}^{\in tb} \!\sum_{{\mathbf q} {\mathbf k}} \!
\frac{f_{\beta\!{\mathbf k}\!+\!{\mathbf q}}\!\!-\!\!f_{\alpha\!{\mathbf k}}} 
{ \epsilon_{\beta\!{\mathbf k}\!+\!{\mathbf q}}\!\! - \!\!\epsilon_{\alpha\!{\mathbf k}} } 
\!\Bigl(\!U^{\dagger ({\bf k})}_{m0,\alpha}\!\Bigr)^{\ast}
\!U^{\dagger ({\bf k}\!+\!{\bf q})}_{n0,\beta} 
\!\Bigl(\!U^{\dagger ({\bf k}\!+\!{\bf q})}_{o0,\beta}\!\Bigr)^{\ast}
\!U^{\dagger ({\bf k})}_{p0,\alpha} \nonumber \\ 
\end{eqnarray} 
corresponding to the local one-center components of a polarization
matrix in the Wannier orbital basis. 
Now, by identifying $\chi_0^t$ in Eq.~(\ref{Wbar2}) as $\chi_0^{{\rm imp}}$ and $\bar{W}$ as $\UDMFT1$, we write the Dyson equation for the effective interaction as 
\begin{eqnarray}
\label{DMFTeq16}
W({\mathbf r}, \!{\mathbf r}^{\prime}\!) \! =\! \UDMFT1\!({\mathbf r}, \!{\mathbf r}^{\prime}\!)
\!+\!\!\! \int \!\!{\rm d}{\mathbf r}^{\prime\prime} \!\!\!\int\!\! {\rm d}{\mathbf r}^{\prime\prime\prime} && 
\UDMFT1\!({\mathbf r}, \!{\mathbf r}^{\prime\prime}\!)
\chi_{0}^{\rm im\!p}\!({\mathbf r}^{\prime\prime}\!, {\mathbf r}^{\prime\prime\prime}\!)
\nonumber \\ && \times W({\mathbf r}^{\prime\prime\prime}\!, {\mathbf r}^{\prime}\!).
\end{eqnarray}
Multiplying this equation by
$\phi^{\ast}_{m0{\mathbf 0}}({\bf r}) \phi_{n0{\mathbf 0}} ({\bf r})
\phi^{\ast}_{o0{\mathbf 0}}({\bf r}^{\prime})$ $\times\phi_{p0{\mathbf 0}}({\bf r}^{\prime})$ and integrating over ${\mathbf r}$ and ${\mathbf r}'$, we have 
\begin{eqnarray}
\label{DMFTeq17}
W_{\mu\nu} = \UDMFT1_{\mu\nu}
+ \sum_{\mu'\nu'}\UDMFT1_{\mu\mu'}C_{\mu'\nu'}
W_{\nu'\nu},
\end{eqnarray}
where we introduce a composite index 
$(\!\mu, \nu\!)$=$\bigl\{\!(mn),(op)\!\bigr\}$
and the matrix element of ${\cal O}$=$\{\!W,\UDMFT1\!\}$ is given by  
\begin{eqnarray}
\label{DMFTeq18}
{\cal O}_{m\!n\!o\!p} \!=\!\! 
\int \! {\rm d}{\mathbf r} \!\int \!{\rm d}{\mathbf r}' 
\phi_{m0{\mathbf 0}}^{\ast}({\mathbf r})
\phi_{n0{\mathbf 0}}({\mathbf r})
{\cal O}({\mathbf r}, {\mathbf r}^{\prime})
\phi_{o0{\mathbf 0}}^{\ast}({\mathbf r}')
\phi_{p0{\mathbf 0}}({\mathbf r}'). \nonumber
\end{eqnarray}
Thus, Eq.~(\ref{DMFTeq17}) is rewritten in a matrix form as 
\label{DMFTeq20dash}
\begin{eqnarray}
\label{DMFTeq20}
\BUDMFT1={\mathbf W}({\mathbf 1}+{\mathbf C} {\mathbf W})^{-1}.
\end{eqnarray}
The equation resembles the unscreening equation (\ref{Wbar2}), but it is formulated entirely in terms of ``local'' one-center quantities, that can be evaluated straightforwardly, allowing for a computationally efficient treatment.

{\it -Application to the Hubbard model}. 
We first apply this scheme to the derivation of $\UDMFT1$ for the 2D single-band Hubbard model.
This is helpful to get insight into the behavior of $\UDMFT1$ with respect to  changes of the electron filling. The Hubbard Hamiltonian reads
\begin{eqnarray}
\label{DMFTeq28} 
{\cal H} \!=\! -t\!\!\sum_{\langle ij\rangle\sigma}  
c_{i\sigma}^{\dagger} \!c_{j\sigma} 
\!-t' \!\!\!\sum_{\langle\!\langle ij \rangle\!\rangle\sigma}
c_{i\sigma}^{\dagger} \!c_{j\sigma} \!
-\! \mu \!\sum_{i\sigma}n_{i\sigma} \!
+U \! \sum_{i} \!  
n_{i\uparrow}\! n_{i\downarrow}, \nonumber  
\end{eqnarray}
where $c_{i\sigma}^{\dagger}$ ($c_{i\sigma}$) creates (annihilates)
 an electron with spin $\sigma$ at site $i$ and $n_{i\sigma}\equiv c_{i\sigma}^{\dagger}c_{i\sigma}$. $t$ ($t'$) is a transfer integral to the (next-)nearest neighbor sites in the $\langle i,j \rangle$ ($\langle\!\langle i,j \rangle\!\rangle$) sums. $U$(=8$t$) and $\mu$ represent the onsite Coulomb repulsion and chemical potential, respectively. 
Taking into account the contributions from the charge susceptibility only
(hence being in accordance with {\it ab initio} methods),
the unscreening equation corresponding to Eq.~(\ref{DMFTeq20})
becomes~\cite{supple} 
\begin{eqnarray}
\label{model_un}
\UDMFT1 = \frac{U}{2} + \frac{\tilde{W}}{1-A\tilde{W}}.
\end{eqnarray}
Here
 $\tilde{W}$ is a diagonal element of a real-space $N$$\times$$N$ matrix
$\tilde{{\mathbf W}}$=$({\mathbf 1}$$-$$\tilde{{\mathbf U}}\mbox{\boldmath
$\chi$}_{0})^{-1}\tilde{{\mathbf U}}$, $\tilde{\mathbf U}$ a diagonal
matrix with elements $\tilde{U}$=$U/2$ and $-A$ (with $A>0$) the diagonal element of the real-space polarization matrix $\bchi1$ with elements
$(\bchi1)_{ij}$=$\chi_0({\mathbf R}_i\!-\!{\mathbf R}_j)$. The latter is
obtained by the
Fourier transform of the reciprocal-space static polarization function
\begin{eqnarray}
\label{DMFTeq29}
\chi_{0}({\mathbf q}) = \frac{2}{N} {\sum_{{\mathbf k}}}
\frac{f_{{\mathbf k}+{\mathbf q}} - f_{{\mathbf k}} } { \xi_{{\mathbf k}+{\mathbf q}} - \xi_{{\mathbf k}}}.
\end{eqnarray}
with $\xi_{{\mathbf k}}$=$-2t(\cos k_x$+$\cos k_y)$$-$$4t'\cos k_x \cos k_y$$-$$\mu$ and $f_{\mathbf k}$ being the eigenvalue and the Fermi distribution function, respectively. 
\begin{figure}[htbp]
\vspace{0cm}
\begin{center}
\includegraphics[width=0.48\textwidth]{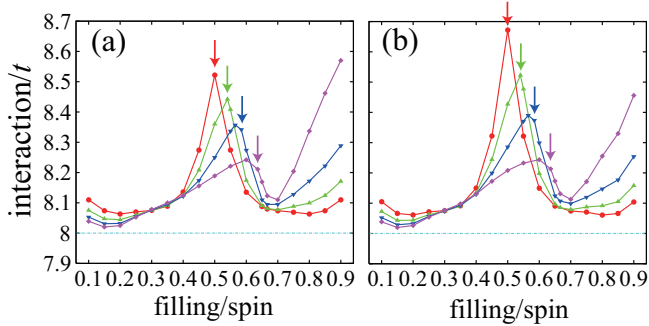}
\caption{(color online) Filling dependence of $\UDMFT1$ calculated (a) with
Eq.~(\ref{model_un}) and (b) with the approximation
[Eq.~(\ref{DMFTeq37})] for $t'=0$ (red), $0.1t$ (green), $0.2t$ (blue), and
$0.3t$ (purple). The arrows indicate the fillings at which the van Hove
singularity resides at the chemical potential.}
\label{fig_U_H}
\end{center}
\end{figure} 

Figure \ref{fig_U_H}(a) shows the filling dependence of $\UDMFT1/t$ with
various $t'$. Contrary to a naive expectation, $\UDMFT1$ is
{\it larger than} $U$. Furthermore, the filling dependence of $\UDMFT1$ is not
monotonic and depends on $t'$. For $t'$=0, $\UDMFT1$ has a strong peak at half
filling where the van Hove singularity resides at the Fermi energy. With
increasing $t'$, the peak shifts to higher filling with reduced peak height,
and another rapid increase emerges at further higher filling.

These filling and $t'$ dependences of $\UDMFT1$ are well understood by
the second-order approximation in \{$B_n$\}~\cite{supple}:
\begin{eqnarray}
\UDMFT1 \sim U + \sum_{n=1}^{N-1}(\tilde{U}B_{n})^{2}\frac{\tilde{U}}{1+\tilde{U}A}, \label{DMFTeq37}
\end{eqnarray}
where
$B_n$$\equiv$$(\bchi1)_{i,i+n}$ is the non-local contribution to the polarization.
Since the second term of the right hand is always positive, the inequality $\UDMFT1$$>$$U$ holds. Figure \ref{fig_U_H}(b) shows the results of $\UDMFT1$ calculated with Eq.~(\ref{DMFTeq37}) for various fillings and $t'$. 
We see in Fig.~\ref{fig_U_H}(b) that Eq.~(\ref{DMFTeq37}) well reproduces the overall trend in Fig.~\ref{fig_U_H}(a).

The inequality $\UDMFT1$$>$$U$ reveals {\it anti-screening} induced by non-local polarizations \{$B_n$\}. This anti-screening is intuitively understood as follows: Suppose that a test charge electron is put on the impurity site. The local polarization screens this electron by creating holes at the impurity site. 
On the other hand, the non-local polarizations induce holes or electrons at other sites. Then, in the second order process, the induced charges create electrons at the impurity site, enhancing the effective repulsion. Since $\frac{\tilde{U}}{1+\tilde{U}A}$ in Eq.~(\ref{DMFTeq37}) varies smoothly with filling \cite{note2}, 
the non-local polarizations \{$B_n$\} indeed dominate the peculiar filling dependence of $\UDMFT1$.


In real materials, off-site Coulomb interactions may play a role. To see this effect, we have studied $\UDMFT1$ for a model with the off-site interaction $1/\epsilon r$ with varying $\epsilon$. We find that the overall filling dependence of $\UDMFT1$ is basically the same as that of the Hubbard model while decreasing $\epsilon$ (i.e., increasing off-site interaction) causes an appreciable reduction of $\UDMFT1$ (not shown). 
The long-range Coulomb interactions connect the onsite polarizations at different sites and thus bring about the screening to the impurity-site interaction. Note that this screening works from the zeroth order in \{$B_n$\}; the approximated $\UDMFT1$ without the contributions from \{$B_n$\} indeed becomes smaller than $U$ and has only  a weak filling dependence. 

{\it -Application to \SrVO3}. 
We next present {\it ab initio} results of ${\mathbf U}^{{\rm D\!M\!F\!T}}$ for \SrVO3. This material is a $d^1$ metal and one of the most benchmarked systems within LDA+DMFT
(local density approximation plus DMFT)~\cite{Rev-Held}. 
On the basis of the DFT band structure, we define
the target bands by the low-energy $t_{2g}$ bands as was done in
Ref.~\cite{cRPA-ex2}. We construct three MLWOs per V site from the $t_{2g}$ Bloch
states and calculate $\UDMFT1$ for these three orbitals. The implementation
details and the convergence checks are elaborated in Ref.~\cite{supple}. 


Table \ref{SrVO3} compares the values of the onsite intra- and inter-orbital Coulomb repulsions ($U$ and $U'$) and Hund's rule coupling ($J$) for the bare (${\mathbf v}$), cRPA (${\mathbf U}^{{\rm cRPA}}$)~\cite{cRPA-ex6}, ${\mathbf U}^{{\rm D\!M\!F\!T}}$, and full-RPA (${\mathbf W}$) interactions. The bare Coulomb interactions ($\sim$15 eV) are largely screened by the high-energy bands, to give $U^{{\rm cRPA}}$$\sim$3 eV. In the present case of SrVO$_3$, $\UDMFT1$ turns out to have a value similar to $U^{\rm cRPA}$. 

\begin{figure}[h]
\vspace{0cm}
\begin{center}
\includegraphics[width=0.48\textwidth]{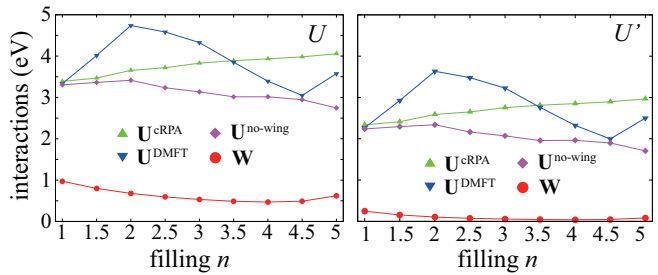}
\caption{(color online)
Filling dependence of intra-orbital (left) and inter-orbital (right) screened Coulomb repulsion of \SrVO3 evaluated within full RPA, cRPA, present scheme ($\BUDMFT1$), and ``no-wing" methods, which are calculated with TAPP~\cite{TAPP}.
}
\label{fig_SrVO3}
\end{center}
\end{figure}

The situation changes drastically, however, when we increase the filling $n$ within the rigid-band approximation. The left and right panels in Fig.~\ref{fig_SrVO3} plot $U$ and $U'$, respectively, against the filling $n$. 
For comparison, we also show the results without the non-local polarizations involving the impurity site, i.e., the interaction parameters calculated without the local one-center and ``wing" components of the polarization matrix in the Wannier basis (``no-wing" method) \cite{note3}.
The result is denoted as ${\mathbf U}^{{\rm no\mathchar`-wing}}$. We see that the filling dependence of $U'$ is similar to that of $U$, except for a constant shift.

As the filling $n$ increases from 1, $\UDMFT1$ increases more rapidly than $U^{{\rm cRPA}}$. This suggests that the non-local anti-screening effect increases more rapidly than the screening. Around $n$$=$2, $\UDMFT1$ turns to decrease, crossing $U^{{\rm cRPA}}$ at $n$$\sim$3.5. Finally around the filling end $n$$\sim$5, $\UDMFT1$ again increases, as seen in the Hubbard model. We see $U^{{\rm no\mathchar`-wing}}$$<$$\UDMFT1$ at all fillings. This is consistent with the model analysis: The non-local contributions to the screening induce an anti-screening and lead to the increase of the onsite interaction.  $U^{{\rm no\mathchar`-wing}}$ is also smaller than $U^{{\rm cRPA}}$ and only weakly depends on the filling, consistently with the model analysis where the off-site Coulomb interaction induces a screening weakly dependent on filling. These comparisons clearly show that the non-local polarization is the main source of the exotic filling dependence of $\UDMFT1$.

It becomes now clear that the similar values of $\UDMFT1$ and $U^{\rm cRPA}$ for SrVO$_3$ is just a consequence of an approximate cancellation of the anti-screening by the non-local polarizations with the screening by the long-range interaction.  In addition, $U^{\rm cRPA}$$\sim$$\UDMFT1$$\sim$$U^{\rm no\mathchar`-wing}$ for SrVO$_3$ is partly ascribed to the small filling of the $d^1$ system where the polarization and screening are not large.

In the previous DMFT studies for the {\it ab initio} model, rather large values of $U$ compared to $U^{\rm cRPA}$ have been needed to reproduce the experimental results (e.g., the insulating behavior of LaTiO$_3$~\cite{Pavarini}). Similarly, for the 2D Hubbard model, the Mott transition takes place at a substantially larger $U$ in the single-site DMFT than in its cluster extension~\cite{Zhang}. These aspects are ascribed to the intersite correlation effects ignored in the single-site DMFT with original $U^{\rm cRPA}$ or $U$. The present scheme with $\UDMFT1$ at least partially takes account of the off-site effects and will improve the results of the DMFT. 
The vertex corrections ignored in the RPA form have been estimated to be small for the conventional cRPA~\cite{MACE}. For the present case, this estimate is left for future studies. 

\begin{table}[h] 
\caption{Onsite bare (${\mathbf v}$), cRPA (${\mathbf U}^{{\rm cRPA}}$),
present-scheme ($\BUDMFT1$), and full-RPA (${\mathbf W}$) interaction
parameters calculated for \SrVO3. The unit of energy is eV. The method was
implemented in two codes, {\it Tokyo Ab initio Program
Package}~\cite{TAPP} (left values) and the {\it Vienna Ab initio Simulation
Package}~\cite{VASP} (right ones), which yield almost identical values
for ${\mathbf U}^{{\rm cRPA}}$. Otherwise, the latter values are
generally 5-10~\% larger than those of the former, since the exact shape
of the orbitals is used in VASP.}
\ 
\centering 
\ 

\begin{tabular}{c@{\ \ \ \  }c@{\ \ \ \ \     }c@{\ \ \ \ \    }c@{\ \ \ \ \   }c} \hline \hline \\ [-5pt]
& ${\mathbf v}$  & \ ${\mathbf U}^{{\rm cRPA}}$ & $\BUDMFT1$ & ${\mathbf W}$ 
\\ \hline \\ [-5pt] 
$U$ & 15.0, 16.0 & 3.39, 3.36 & 3.33, 3.46 & 0.97, 1.12 \\ 
$U'$& 13.7, 14.8 & 2.34, 2.35 & 2.27, 2.47 & 0.25, 0.30 \\ 
$J$ & 0.59, 0.55 & 0.47, 0.49 & 0.47, 0.47 & 0.33, 0.39 \\ 
\hline \hline
\end{tabular} 
\label{SrVO3} 
\end{table}

{\it -Conclusion}.
We have examined a scheme to evaluate the effective onsite 
interaction $\UDMFT1$ for the DMFT. Through the analysis based on the 
Hubbard model, we have found unexpectedly 
an anti-screening effect induced by non-local polarizations, which competes with the screening effects caused by the off-site Coulomb interaction in real materials. The anti-screening causes a non-trivial  
filling dependence of $\UDMFT1$ 
and 
increases the effective interaction.
Combining  the present method with DFT, we have indeed shown that $\UDMFT1$ for \SrVO3 
exhibits
 non-trivial filling dependence if the chemical potential is varied.

\begin{acknowledgements}
{\it -Acknowledgments}.
We would like to thank Takashi Miyake for fruitful discussions. This work was supported by Grants-in-Aid for Scientific Research (No.~22740215, 22104010, 23110708, 23340095, 22340090) from MEXT and JST-PRESTO, Japan
and the Austrian Science fund through F41 (SFB ViCoM)  and I597-16.
\end{acknowledgements}

%% file: supplement.tex
\newcommand{\uu}[1]{\mathbf{#1}_{_{\uparrow\uparrow}}}
\newcommand{\ud}[1]{\mathbf{#1}_{_{\uparrow\downarrow}}} 
\newcommand{\du}[1]{\mathbf{#1}_{_{\downarrow\uparrow}}}
\newcommand{\dd}[1]{\mathbf{#1}_{_{\downarrow\downarrow}}} 
\arraycolsep=0.0em
\def\UH1{U^{{\rm lat}}}
\def\UDMFT1{U^{\rm{D\!M\!F\!T}}}
\def\UcDMFT1{\tilde{U}^{\rm{D\!M\!F\!T}}}
\def\UDMFTh1{U^{\rm{D\!M\!F\!T\mathchar`-high}}}
\def\UDMFTl1{U^{\rm{D\!M\!F\!T\mathchar`-low}}}
\def\BUDMFT1{{\mathbf U}^{\rm{D\!M\!F\!T}}}
\def\BUDMFTh1{{\mathbf U}^{\rm{D\!M\!F\!T\mathchar`-high}}}
\def\BUDMFTl1{{\mathbf U}^{\rm{D\!M\!F\!T\mathchar`-low}}}
\def\Zij1{Z_{\parallel \ \!\!\! i-j \ \!\!\! \parallel}}
\def\ij1{\parallel \ \!\!\! i-j \ \!\!\! \parallel}
\def\tij1{t_{ij}^{mn}}
\def\Uij1{U_{ij}^{mnop}}
\def\gs0ij{{{\mathbf G}^{0}_{ij,\sigma}}}
\def\Gs1ij{{{\mathbf G}_{ij,\sigma}}}
\def\g0ij{{{\mathbf G}^{0}_{ij}}}
\def\G1ij{{{\mathbf G}_{ij}}}
\def\Utij1{{{\tilde{U}}_{ij}^{mnop}}}
\def\Wh1{W^{\rm high}}
\def\Wl1{W^{\rm low}}
\def\Us1{U^{\rm scr}}
\def\boldchi1{\mbox{\boldmath $\chi$}_{0}}
\def\bchi1{\mbox{\boldmath $\chi$}_{0}'}
\def\chib2{\mbox{\boldmath $\chi$}}
\def\chipp3{\mbox{\boldmath $\chi$}_{0}''}
\def\SrVO3{SrVO$_3$}
\def\mnop1{m\!n\!o\!p}
\def\ga{{\bf g}_1}
\def\gb{{\bf g}_2}
\def\gc{{\bf g}_3}
\def\gd{{\bf g}_4}
\def\gi{{\bf g}_i}

{\LARGE Supplemental  Materials}

\affiliation{%
Department of Applied Physics, University of Tokyo, and JST CREST,
7-3-1 Hongo, Bunkyo-ku, Tokyo, 113-8656, Japan}

\renewcommand{\theequation}{S.\arabic{equation}}
\setcounter{equation}{0}
\renewcommand{\tablename}{Table S}

\noindent

\section{ S.1 DERIVATION OF EQS.~(12) and (14)}
An RPA fully-screened
interaction $W$ may be expressed as
\begin{eqnarray}
\label{DMFTeq31}
\overline{\mathbf{W}} = 
\overline{\mbox{{\boldmath $\epsilon$}}}^{-1}\overline{\mathbf{U}},\quad
 \overline{\mbox{{\boldmath $\epsilon$}}}={{\mathbf 1}}\! - \! \overline{{\mathbf U}}\overline{\mbox{\boldmath
$\chi$}}_{0}.
\end{eqnarray}   
Here
$\overline{\mathbf{X}}=\left[\overline{\mathbf{W}},\overline{\mathbf{U}},\overline{\mbox{{\boldmath $\chi$}}}_0,\overline{\mbox{{\boldmath $\epsilon$}}}\right]$
are $2N$$\times$$2N$ matrices decomposed into their spin channels
according to 
\begin{eqnarray}
\label{spin_decomposition} 
\overline{\mathbf{X}}=\left(
\begin{array}{cc}
\uu{X} & \ud{X}\\
\du{X} & \dd{X}\\
\end{array}
\right).
\end{eqnarray}
With this decomposition, $\overline{{\mathbf U}}$ and $\overline{\mbox{\boldmath $\chi$}}_{0}$ are written as 
\begin{eqnarray}
\label{DMFTeq34} 
\overline{{\mathbf U}}=\left(
\begin{array}{cc}
 {\mathbf 0} & \ {\mathbf U} \\
 {\mathbf U} & \ {\mathbf 0} \\
\end{array}
\right)  
\ \ {\rm and}\ \ 
\overline{\mbox{\boldmath $\chi$}}_{0}=
\left(
\begin{array}{cc}
 \frac{1}{2}\boldchi1 & \ {\mathbf 0}       \\
 {\mathbf 0}          & \ \frac{1}{2}\boldchi1 \\
\end{array}
\right),
\end{eqnarray}
respectively, 
where ${\mathbf U}$ is a
diagonal matrix with elements $U$ and $\boldchi1$ is a real-space polarization
matrix.
In the Hubbard model, only the onsite components
$W=\left[ \ud{W} \right]_{ii}$ are relevant, which are given by  
\begin{equation}\label{W_explicit}
\begin{split}
W&=\left[ \uu{\epsilon}^{-1}\ud{U}+\ud{\epsilon}^{-1}\dd{U} \right]_{ii}\\
 &=\left[\uu{\epsilon}^{-1}\mathbf{U}\right]_{ii}.
\end{split}
\end{equation}
According to Eqs. (5.9) and (5.10) in Ref.~\cite{Cohen}, the inverse dielectric
matrix in the $\uparrow\uparrow$ spin channel is
$(\mathbf{1}-\left(\frac{1}{2}\mathbf{U}\mbox{{\boldmath $\chi$}}_0\right)^2)^{-1}$
so that 
we obtain  
\begin{eqnarray}
\label{ap_eq1.5}
W = \biggl[\Bigl(\mathbf{1}\!-\!\bigl(\frac{1}{2}{\mathbf U}\boldchi1\bigr)^2\Bigr)^{-1} 
{\mathbf U}  \biggr] _{ii}.
\end{eqnarray} 
This equation is 
also written as
\begin{eqnarray}
\label{ap_eq1.6}
W = \biggl[{\mathbf U} + \frac{1}{2} {\mathbf U}(\chib2_{C}-\chib2_{S}){\mathbf U} \biggr]_{ii}
\end{eqnarray} 
with $\chib2_{C}$ ($\chib2_{S}$) being the charge (spin) susceptibility given by 
$\chib2_{C}$=$({\mathbf 1}$$-$$\frac{1}{2}\boldchi1{\mathbf U})^{-1} \frac{1}{2}\boldchi1$ [$\chib2_{S}$=$({\mathbf 1}$+$\frac{1}{2}\boldchi1{\mathbf U})^{-1} \frac{1}{2}\boldchi1$]. 
In line with {\it ab initio} methods, which only take charge fluctuations
into account, we consider the term related to $\chib2_{C}$
only; the resulting expression for $W$ is 
\begin{eqnarray}
\label{ap_eq5}
W 
&=& 
\biggl[{\mathbf U} + \frac{1}{2} {\mathbf U}\chib2_{C}{\mathbf U} \biggr]_{ii} 
\nonumber \\ 
&=& 
\biggl[{\mathbf U} + \frac{1}{2} {\mathbf U}\ \biggl({\mathbf 1} -\frac{1}{2}\boldchi1{\mathbf U}\biggr )^{-1}\ \frac{1}{2}\boldchi1{\mathbf U} \biggr]_{ii} 
\nonumber \\
&=& 
\biggl[{\mathbf U} + \tilde{{\mathbf U}} \bigl({\mathbf 1} - \boldchi1\tilde{{\mathbf U}}\bigr )^{-1} \boldchi1\tilde{{\mathbf U}} \biggr]_{ii} 
\nonumber \\ 
&=& 
\biggl[\tilde{{\mathbf U}}  +  \bigl({\mathbf 1}+\tilde{{\mathbf U}}\boldchi1 + (\tilde{{\mathbf U}}\boldchi1)^2 + \cdots \bigr) \tilde{{\mathbf U}} \biggr]_{ii} 
\nonumber \\ 
&=& \biggl[\tilde{{\mathbf U}} + \bigl({\mathbf 1} -\tilde{{\mathbf U}}\boldchi1 \bigr )^{-1}\tilde{{\mathbf U}}\biggr]_{ii} 
\nonumber \\ 
&=& 
\ \ \tilde{U} + \tilde{W}, 
\end{eqnarray} 
where $\tilde{U}$=$U/2$ and $\tilde{W}$=$\bigr[\bigl({\mathbf
1}$$-$$\tilde{{\mathbf U}}\boldchi1\bigr)^{-1}\tilde{{\mathbf U}}\bigr]_{ii}$. 

We now decompose the total polarization $\boldchi1$ into the two parts,
\begin{eqnarray}
\label{ap_eq5.0}
\boldchi1^{t} = 
\left(
\begin{array}{cc}
     -A     & \ {\mathbf 0} \\
{\mathbf 0} & \ {\mathbf 0} \\
\end{array}
\right) 
\ \ {\rm and}\ \ 
\bchi1 = 
\left(
\begin{array}{cc}
     0        & \ {\mathbf B}^{\rm T} \\
 {\mathbf B}  & \      \bchi1'        \\
\end{array}
\right), 
\end{eqnarray}
where ${\mathbf B}$=$(B_{1}, B_{2},\ \! \cdots, B_{N-1})^{\rm T}$ and $\bchi1'$ is an $(N-1)$$\times$$(N-1)$ matrix.
Then, replacing $\boldchi1$ with $\bchi1$ in Eqs.~(\ref{DMFTeq31}-7), we obtain
\begin{eqnarray} 
\label{s2_eq21}
 \UDMFT1 = \tilde{U} + \UcDMFT1 
\end{eqnarray} 
with
\begin{eqnarray}
\label{s2_eq2}
\UcDMFT1 = \biggl[ \bigl({\mathbf 1}-\tilde{{\mathbf U}}\bchi1 \bigr)^{-1}\tilde{{\mathbf U}}\biggr]_{11}.  
\end{eqnarray}
The above
 derivation of $\UDMFT1$ is based on the screening approach 
of
 Eq.~(2). 
On the other hand, $\UcDMFT1$ can also be obtained 
in the unscreening approach of Eq.~(3) as
\begin{eqnarray}
\label{s2_eq22}
\UcDMFT1 
= 
\biggl[\tilde{{\mathbf W}} \bigl({\mathbf 1} + \boldchi1^{t} \tilde{{\mathbf W}} \bigr)^{-1} \biggr]_{11}
= \frac{\tilde{W}}{1-A\tilde{W}}. 
\end{eqnarray}
Eqs.~(\ref{s2_eq21}) and (\ref{s2_eq22}) give Eq.~(12) in the main text.
Again using Eqs.~(5.9) and (5.10) in Ref.~\cite{Cohen}, Eq.~(\ref{s2_eq2}) is further recast into
\begin{eqnarray}
\label{ap_eq5.1}
\UcDMFT1 = \frac{1}{1-\tilde{U}^2 {\mathbf B}^{\rm T} ({\mathbf 1} - \tilde{U} \chipp3)^{-1}{\mathbf B} }\tilde{U}.
\end{eqnarray}
Hence,
 up to the second order in \{$B_n$\}, we obtain 
\begin{eqnarray}
\label{ap_eq5.2}
\UDMFT1 \sim \tilde{U} +   \biggl(  1 + \frac{  \tilde{U}^2 {\mathbf B}^{\rm T} {\mathbf B}  } { 1 + \tilde{U} A} \biggr)\tilde{U},
\end{eqnarray}
which is
 equivalent to Eq.~(14).

\noindent
\section{ S.2 IMPLEMENTATION DETAILS OF ${\mathbf U}^{{\rm D\!M\!F\!T}}$ IN THE PLANE-WAVE BASIS-SET CODE AND COMPUTATIONAL RESULTS}

Here, we describe implementation details for the {\it ab initio} ${\mathbf
U}^{{\rm D\!M\!F\!T}}$ calculations. The calculation is performed with the
norm-conserving pseudopotential and plane-wave basis set and the projector
augmented wave method, respectively~\cite{paw1,paw2}. In the plane-wave
basis-set calculation, two different cutoffs for the plane waves are
conventionally used; the low-momentum cutoff ${\rm g}_{\rm cut}^{\rm low}$ for
the polarization function and the high-momentum cutoff ${\rm g}_{\rm cut}^{\rm
high}$ for orbitals. In general, the structure of the polarization function in
real space is smooth compared to that of the wavefunction, so 
we can employ the smaller cutoff 
and it considerably reduces the computational cost.
In the ${\mathbf U}^{{\rm D\!M\!F\!T}}$ calculation in Eq.~(11) in the main text, however, we should be careful about the use of the two different cutoffs. 

The Dyson equation Eq.~(10) is written in the momentum space with the double Fourier transform~\cite{Hybertsen} as 
\begin{eqnarray}
\label{Wlow}
W\!_{{\ga\gb}}\!&=&\!U\!_{{\ga\gb}}^{{\rm D\!M\!F\!T}}\!\!\!+\!\!\!\sum_{{\gc\gd}}\!U\!_{{\ga\gc}}^{{\rm D\!M\!F\!T}}\!\chi\!_{{\gc\gd}}^{{\rm im\!p}}\!W\!_{{\gd\gb}} ({|\gi|}\!\!<\!\!{\rm g}_{\rm cut}^{\rm low}\!) \\
W\!_{{\ga\gb}}&=&U\!_{{\ga\gb}}^{{\rm D\!M\!F\!T}}\!=\!v_{{\ga}}\delta_{{\ga\gb}}\ 
({\rm g}_{cut}^{{\rm low}}\!\!\le\!\!|\gi|\!\!\le\!\!{\rm g}_{\rm cut}^{\rm high}\!),
\label{Whigh}
\end{eqnarray}
where $\ga$-$\gd$ are reciprocal wave vectors associated with the superlattice~\cite{note} and
$v_{{\bf g}}$$=$$4\pi/|{\bf g}|^2$ is the Fourier transform of the bare
Coulomb interaction $v$. In Eq.~(\ref{Whigh}) we have used the fact that $\chi_{{\bf
gg'}}^{{\rm imp}}$ vanishes outside ${\rm g}_{\rm cut}^{\rm low}$.

Recognizing this aspect, we define the low- and high-momentum contributions for $W_{\mu\nu}$, defined in Eq.~(10), as  
\begin{eqnarray}
\label{DMFTeq22}
W\!_{\mu\nu}^{{\rm low}}\!&=&\!\frac{1}{V}\!\!\sum_{{\bf gg'}}^{{\rm low}}\!\langle \phi_{m\!0{\bf 0}}\!|e^{i{\bf gr}}|\!\phi_{n\!0{\bf 0}}\!\rangle\!W\!\!_{{\bf gg'}}\!\langle \phi_{o\!0{\bf 0}}\!|e^{-i{\bf g'r'}}\!|\!\phi_{p\!0{\bf 0}}\!\rangle\!, \\
\label{DMFTeq23}
W\!_{\mu\nu}^{{\rm high}}\!&=&\!\frac{1}{V}\!\!\sum_{{\bf g}}^{{\rm high}}\!\langle \phi_{m0{\bf 0}}\!|e^{i{\bf gr}}|\!\phi_{n0{\bf 0}}\rangle v_{{\bf g}} \langle \phi_{o0{\bf 0}}\!|e^{-i{\bf gr'}}\!|\!\phi_{p0{\bf 0}}\!\rangle.
\end{eqnarray} 
Here, $V$ is the crystal volume and $W_{\mu\nu}$=$W_{\mu\nu}^{{\rm
low}}$+$W_{\mu\nu}^{{\rm high}}$. The sum in Eq.~(\ref{Wlow}) is taken for the
reciprocal vector within ${\rm g}_{cut}^{{\rm low}}$, while the sum in
Eq.~(\ref{Whigh}) runs over the reciprocal vector for ${\rm g}_{cut}^{{\rm
low}}$$\le$$|{\bf g}|$$\le$${\rm g}_{cut}^{{\rm high}}$. Similarly,
$U_{\mu\nu}^{{\rm D\!M\!F\!T}}$ is written as the sum of $U_{\mu\nu}^{{\rm
D\!M\!F\!T\mathchar`-low}}$ and $U_{\mu\nu}^{{\rm
D\!M\!F\!T\mathchar`-high}}$. Inserting Eq.~(\ref{Wlow}) into
Eq.~(\ref{DMFTeq22}) with the double Fourier transform of $\chi_{0}^{{\rm
imp}}$, we obtain
\begin{eqnarray}
W_{\mu\nu}^{{\rm low}} = U_{\mu\nu}^{{\rm D\!M\!F\!T\mathchar`-low}} + \sum_{\mu'\nu'} U_{\mu\mu'}^{{\rm D\!M\!F\!T\mathchar`-low}} C_{\mu'\nu'} W_{\nu'\nu}^{{\rm low}} 
\end{eqnarray}
or in the matrix form 
\begin{eqnarray}
\label{DMFTeq25.5}
{\mathbf W}^{\rm low}=\BUDMFTl1 + \BUDMFTl1 {\mathbf C}{\mathbf W}^{\rm low}.
\end{eqnarray}
Since ${\mathbf U}^{{\rm D\!M\!F\!T}}$=${\mathbf U}^{{\rm D\!M\!F\!T\mathchar`-low}}$+${\mathbf U}^{{\rm D\!M\!F\!T\mathchar`-high}}$, after some manipulations, we obtain 
\begin{equation}
\BUDMFT1={\mathbf W}^{\rm low}({\mathbf 1}+{\mathbf C} {\mathbf W}^{\rm low})^{-1}  + {\mathbf V}^{\rm high}
\label{DMFTeq27}
\end{equation}
with ${\mathbf V}^{\rm high}$ (=${\mathbf W}^{\rm high}$=$\BUDMFTh1$) being the matrix of $v$ at high momenta Eq. (\ref{DMFTeq23}). In the actual calculation, this expression is used.

As a note on the numerical calculation, we remark some details for calculating
the
polarization function in a metallic system.
The target-band polarization $\chi_0^{tb}({\bf r},{\bf r'})$ in Eq.~(4) in the
main text is given in the momentum space with the double Fourier transform as

\begin{eqnarray}
\chi^{tb}_{{\bf G} {\bf G}'}({\bf q})= 2
\sum_{{\bf k}} \sum_{\alpha \beta}^{\in tb} &&  
\frac{ f_{\beta {\bf k}+{\bf q}}\!-\!f_{\alpha {\bf k}}}
{\epsilon_{\beta {\bf k}+{\bf q}}\!-\!\epsilon_{\alpha {\bf k}}} 
\langle \psi_{\alpha {\bf k}} 
| e^{-i ({\bf q}+{\bf G}) \cdot {\bf r}} | 
\psi_{\beta {\bf k}+{\bf q}} \rangle \nonumber \\ 
&&\times
\langle \psi_{\beta {\bf k}+{\bf q}} 
| e^{i ({\bf q}+{\bf G}') \cdot {\bf r}'} 
| \psi_{\alpha {\bf k}} \rangle.
\label{eq:chi} 
\end{eqnarray}
Here, ${\mathbf G}$ is a reciprocal lattice vector for the primitive lattice
and ${\mathbf q}$ is a wave vector in the first Brillouin zone.
$\{\psi_{\alpha {\bf k}} \}$, $\{ \epsilon_{\alpha {\bf k}} \}$, and $\{ f_{\alpha {\bf k}} \}$ 
are the Bloch states, their energies, and occupancies, respectively, and the
band summation runs over the target bands only. In the calculation of
$\chi^{tb}_{{\bf G} {\bf G}'}({\bf q})$ of the metallic system, the ${\mathbf
k}$ integral on the right hand must be performed carefully, because the
expression includes a  numerical instability due to the Lindhard part. To avoid
the instability, we use the Wannier interpolation scheme~\cite{WI}; 
we interpolate the original ${\bf k}$-point data (of about
10$\times$10$\times$10) for the eigenvalues \{$\epsilon_{\alpha {\bf k}}$\} and
interstate matrix elements \{$\langle \psi_{\beta {\bf k}+{\bf q}}|e^{i ({\bf q}+{\bf
G}) {\bf r}}|\psi_{\alpha {\bf k}} \rangle$\}, to obtain the data on a denser
${\bf k}$ grid (about 30$\times$30$\times$30).
After such an interpolation, the ${\bf k}$ integration is performed with the
generalized tetrahedron method~\cite{tetra} to obtain both, real and imaginary parts of $\chi^{tb}_{{\bf G} {\bf G}'}({\bf q})$.

We also need a careful treatment of poles at $\epsilon_{\beta {\bf k+q}}\!=\!\epsilon_{\alpha {\bf k}}$ in Eq.~(\ref{eq:chi}), for which we rewrite 
\begin{eqnarray}
\label{switch} 
\frac{ f_{\beta {\bf k}+{\bf q}} - f_{\alpha {\bf k}}} 
{\epsilon_{\beta {\bf k}+{\bf q}} - \epsilon_{\alpha {\bf k}}}
\sim \delta 
\Bigl( 
\frac{\epsilon_{\beta {\bf k}+{\bf q}} + \epsilon_{\alpha {\bf k}}}{2} 
- \epsilon_{{\rm F}} 
\Bigr),
\end{eqnarray}
Based on the central-difference approximation of the Fermi-Dirac function with
The Fermi level $\epsilon_{{\rm F}}$. 
Switching to the $\delta$ function in
Eq.~(\ref{switch}) is performed in the threshold $|\epsilon_{\beta {\bf
k}+{\bf q}}-\epsilon_{\alpha {\bf k}}|$$<$0.06 eV and the $\delta$ function is
treated with a smearing factor of 0.03 eV. With the resulting target-band
polarization $\chi^{tb}$ and the rest polarization $\chi^{r}$~\cite{cRPA-exS6},
the fully screened RPA Coulomb interaction ${\mathbf W}^{{\rm low}}$ in Eq.~(\ref{DMFTeq27}) is calculated, where 
the $W_{{\mathbf G}{\mathbf G}'}({\mathbf q})$ interaction at ${\mathbf
q}\to{\mathbf 0}$ limit is treated following Ref.~\cite{Cohen}. 
The same treatment is applied to the evaluation of the Wannier matrix elements of $C_{m\!n\!o\!p}$ in Eq.~(8) \cite{noteS2}. With all these treatments, the present ${\mathbf U}^{{\rm D\!M\!F\!T}}$ calculation ensures the accuracy within several percent. 

If not otherwise noted, the density-functional theory calculations for
SrVO$_3$ were performed with {\it Tokyo Ab initio Program
Package}~\cite{TAPPS}, which is based on the pseudopotential plus plane-wave
framework. The exchange-correlation functional is calculated within the
generalized-gradient approximation with Perdew-Burke-Ernzerhof (PBE)
parameterization~\cite{PBE}, and the Troullier-Martins norm-conserving
pseudopotentials~\cite{TM} in the Kleinman-Bylander representation~\cite{KB}
is adopted. In the present calculation for Fig.~2 in the main text, the cutoff
energies for wavefunctions and polarization functions are set to 49 Ry and 25
Ry, respectively, and we employ 11$\times$11$\times$11 ${\mathbf k}$ points.
The Brillouin-zone integrals are evaluated using the generalized
tetrahedron method~\cite{tetra} after interpolation to a
33$\times$33$\times$33 ${\mathbf k}$ mesh. 

Where noted, additional calculations were performed using the {\it Vienna Ab initio Simulation Package} (VASP), using projector augmented
waves and the local density approximation. The plane wave  cutoff energies for the orbitals and response functions were set to  
$414~\mathrm{eV}$ ($30~\mathrm{Ry}$) and 250~eV ($18~$Ry), respectively.
Extrapolation to a high energy cutoff (500~eV) was performed using Eq. (\ref{DMFTeq27}). In VASP no intermediate extrapolation
to a denser k-point grid was performed. Instead, in Eq. (\ref{eq:chi}), the Fermi occupancy function $f(\epsilon)$ was
replaced by a Methfessel Paxton smearing function with $\sigma=0.1$ \cite{MP}, 
and consistent with metallic screening 
$\mathbf{W}_{{\bf 0} {\bf 0}}({\bf q}\to 0)$ was set to 0.

Figure S~\ref{fig_band} shows our calculated band structure of \SrVO3 (a) and
the density of states for the $t_{2g}$ bands (b). The arrows in the panel (b)
indicate the Fermi levels for the fillings $n$=1.0 to 5.0 with the interval
0.5. We see that the van Hove singularity nearly corresponds to the Fermi
level at the filling $n=4.0$.
\renewcommand{\figurename}{Fig. S}
\setcounter{figure}{0}
\begin{figure}[htbp]
\vspace{0cm}
\begin{center}
\includegraphics[width=0.48\textwidth]{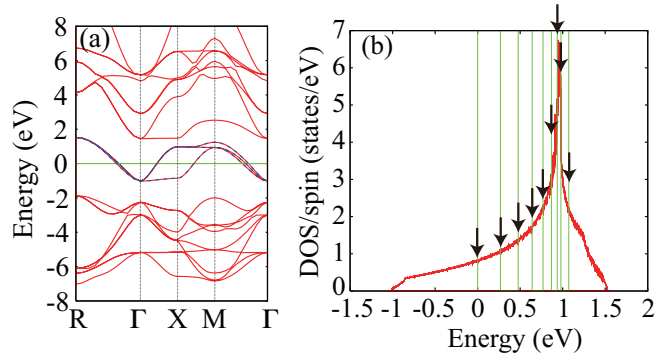}
\caption{(color online) (a)  Calculated electronic band structure of \SrVO3. The interpolated band dispersions for the $t_{2g}$ bands are depicted as blue dashed lines, which cross the Fermi level. (b) Calculated density of states for the $t_{2g}$ bands. Black arrows indicate the Fermi level for the filling $n=1.0$-$5.0$ from left to right for the values shown in Table~S~\ref{U-n}} . 
\label{fig_band}
\end{center}
\end{figure} 

We show in Table~S~\ref{CV-k} and S~\ref{vasp_results} the convergence behavior 
of ${\mathbf U}^{{\rm D\!M\!F\!T}}$ calculated for SrVO$_3$ against the sampling ${\mathbf k}$ points
using the  {\it Tokyo Ab initio Program Package}. The table lists the values for the onsite intra- and inter-orbital Coulomb repulsions ($U$ and $U'$) and Hund's rule coupling ($J$). The usual constrained random-phase-approximation (cRPA) (${\mathbf U}^{{\rm cRPA}}$)~\cite{cRPA-exS6} and full-RPA (${\mathbf W}$) results are also shown for comparison. 
We see that the results are almost converged 
at 6$\times$6$\times$6 or 7$\times$7$\times$7 ${\mathbf k}$-point samplings. 
Despite a less sophisticated interpolation procedure the results using the
{\it Vienna Ab initio Simulation Package} (VASP) show a very similar convergence behavior. Again the error is reduced to few percent
at 7$\times$7$\times$7 ${\mathbf k}$-points, although a sizeable scattering prevails in both codes.

Table~S~\ref{CV-g} 
shows the convergence behavior 
against the cutoff momentum ${\rm g}_{\rm cut}^{\rm low}$ for the polarization function. We see that the convergence is attained around ${\rm g}_{\rm cut}^{\rm low}$$\sim$25 Ry. Finally, Table~S~\ref{U-n} lists
the interaction parameters calculated at the fillings $n$=1.0-5.0,
 which are used for the plot in Fig.~2 in the main text. In this table, we add the ``no-wing" data (${\mathbf U}^{{\rm no\mathchar`-wing}}$). For the definition of ${\mathbf U}^{{\rm no\mathchar`-wing}}$, see the main text.

\begin{table*}
\setcounter{table}{0}
\caption[t]{Convergence behavior of ${\mathbf U}^{{\rm cRPA}}$, ${\mathbf U}^{{\rm D\!M\!F\!T}}$, and ${\mathbf W}$ to the sampling $k$ points of SrVO$_3$ for the  {\it Tokyo Ab initio Program Package}. The cutoff energy for polarization function is 25 Ry.}
\begin{center}
\begin{tabular}{rccccccccccc}
\hline \hline \\ [-5pt]
 & \multicolumn{3}{c}{${\mathbf U}^{{\rm cRPA}}$} & 
 & \multicolumn{3}{c}{${\mathbf U}^{{\rm D\!M\!F\!T}}$} & 
 & \multicolumn{3}{c}{${\mathbf W}$} \\ 
 \hline
 & $U$ & $U'$ & $J$ & 
 & $U$ & $U'$ & $J$ & 
 & $U$ & $U'$ & $J$ \\ 
 \hline 
5$\times$5$\times$5  \ \  
            & 3.40  &   2.34    &  0.47   & \ \ 
            & 3.48  &   2.41    &  0.41   & \ \ 
            & 0.93  &   0.23    &  0.33   \\
6$\times$6$\times$6  \ \  
            & 3.50  &   2.45    &  0.47   & \ \ 
            & 3.44  &   2.37    &  0.47   & \ \ 
            & 0.98  &   0.25   &   0.33   \\
7$\times$7$\times$7  \ \  
            & 3.42  &   2.37    &  0.47   & \ \ 
            & 3.37  &   2.30    &  0.47   & \ \ 
            & 0.97  &   0.25    &  0.33  \\ 
8$\times$8$\times$8  \ \  
            & 3.32  &   2.27    &  0.47   & \ \ 
            & 3.26  &   2.20    &  0.48   & \ \ 
            & 0.96  &   0.24    &  0.33   \\ 
9$\times$9$\times$9  \ \  
            & 3.27  &   2.22    &  0.47   & \ \ 
            & 3.22  &   2.16    &  0.48   & \ \ 
            & 0.97  &   0.25    &  0.33   \\ 
10$\times$10$\times$10 \ \ 
            & 3.44  &   2.38    &  0.47   & \ \ 
            & 3.39  &   2.33    &  0.47   & \ \ 
            & 0.98  &   0.25    &  0.33   \\ 
11$\times$11$\times$11 \ \ 
            & 3.39  &   2.34    &  0.47   & \ \ 
            & 3.33  &   2.27    &  0.47   & \ \ 
            & 0.97  &   0.25   &  0.33   \\ 
\hline \hline 
\end{tabular}
\end{center}
\label{CV-k}
\end{table*}

\begin{table*}
\caption[t]{Convergence behavior of ${\mathbf U}^{{\rm cRPA}}$, ${\mathbf
U}^{{\rm D\!M\!F\!T}}$, and ${\mathbf W}$ to the sampling $k$ points of
SrVO$_3$ for the  {\it Vienna Ab initio Simulation Package}.}
\begin{center}
\begin{tabular}{rccccccccccc}
\hline \hline \\ [-5pt]
 & \multicolumn{3}{c}{${\mathbf U}^{{\rm cRPA}}$} & 
 & \multicolumn{3}{c}{${\mathbf U}^{{\rm D\!M\!F\!T}}$} & 
 & \multicolumn{3}{c}{${\mathbf W}$} \\ 
 \hline
 & $U$ & $U'$ & $J$ & 
 & $U$ & $U'$ & $J$ & 
 & $U$ & $U'$ & $J$ \\ 
 \hline 
3$\times$3$\times$3  \ \  
            & 3.45  &    2.43    &  0.50   & \ \ 
            & 6.38  &   5.38    &  0.48   & \ \ 
            & 1.02  &   0.23    &  0.38   \\
4$\times$4$\times$4  \ \  
            & 3.31  &   2.30    &  0.49   & \ \ 
            & 5.25  &   4.26    &  0.47   & \ \ 
            & 1.00  &   0.22   &   0.38   \\
5$\times$5$\times$5  \ \  
            & 3.31  &   2.30    &  0.49   & \ \ 
            & 3.94  &   2.95    &  0.47   & \ \ 
            & 1.07  &   0.26   &   0.39   \\
6$\times$6$\times$6  \ \  
            & 3.35  &   2.34    &  0.49   & \ \ 
            & 3.50  &   2.51    &  0.47   & \ \ 
            & 1.11  &   0.29   &   0.39   \\
7$\times$7$\times$7  \ \  
            & 3.38  &   2.36    &  0.49   & \ \ 
            & 3.51  &   2.53    &  0.47   & \ \ 
            & 1.17  &   0.34    &  0.40  \\ 
8$\times$8$\times$8  \ \  
            & 3.36  &   2.35    &  0.49   & \ \ 
            & 3.46  &   2.47    &  0.47   & \ \ 
            & 1.12  &   0.30    &  0.39   \\ 
9$\times$9$\times$9  \ \  
            & -  &   -    &  -   & \ \ 
            & 3.42  &   2.43    &  0.47   & \ \ 
            & 1.10  &   0.29    &  0.39   \\ 
10$\times$10$\times$10 \ \ 
            & -  &   -    &  -   & \ \ 
            & 3.42  &   2.43    &  0.47   & \ \ 
            & 1.11  &   0.30    &  0.39   \\ 
11$\times$11$\times$11 \ \ 
            & -  &   -    &  -   & \ \ 
            & 3.48  &   2.49    &  0.47   & \ \ 
            & 1.14  &   0.31    &  0.39   \\ 
\hline \hline 
\end{tabular}
\end{center}
\label{vasp_results}
\end{table*}


\begin{table*}
\caption[t]{Convergence behavior of ${\mathbf U}^{{\rm cRPA}}$, ${\mathbf U}^{{\rm D\!M\!F\!T}}$, and ${\mathbf W}$ to the cutoff energy for polarization function ${\rm g}_{\rm cut}^{\rm low}$ for the  {\it Tokyo Ab initio Program Package}. The sampling $k$ points are fixed at 7$\times$7$\times$7 and, in the interpolation of the polarization calculation, the 21$\times$21$\times$21 $k$-grid is employed.}
\begin{center}
\begin{tabular}{rccccccccccc}
\hline \hline \\ [-5pt]
 & \multicolumn{3}{c}{${\mathbf U}^{{\rm cRPA}}$} & 
 & \multicolumn{3}{c}{${\mathbf U}^{{\rm D\!M\!F\!T}}$} & 
 & \multicolumn{3}{c}{${\mathbf W}$} \\ 
 \hline
 & $U$ & $U'$ & $J$ & 
 & $U$ & $U'$ & $J$ & 
 & $U$ & $U'$ & $J$ \\ 
 \hline 
 10 Ry  \ \  
            & 3.48  &   2.37    &  0.51   & \ \ 
            & 3.38  &   2.28    &  0.51   & \ \ 
            & 1.22  &   0.26    &  0.45   \\
 15 Ry  \ \  
            & 3.48  &   2.39    &  0.49   & \ \ 
            & 3.39  &   2.30    &  0.49   & \ \ 
            & 1.13  &   0.27    &  0.39   \\
 20 Ry  \ \  
            & 3.44  &   2.38    &  0.48   & \ \ 
            & 3.37  &   2.30    &  0.48   & \ \ 
            & 1.04  &   0.26    &  0.36   \\ 
 25 Ry  \ \  
            & 3.42  &   2.37    &  0.47   & \ \ 
            & 3.37  &   2.30    &  0.47   & \ \ 
            & 0.97  &   0.25    &  0.33   \\ 
 30 Ry  \ \  
            & 3.41  &   2.36    &  0.47   & \ \ 
            & 3.37  &   2.30    &  0.47   & \ \ 
            & 0.94  &   0.24    &  0.32   \\ 
35 Ry  \ \ 
            & 3.40  &   2.36    &  0.47   &  \ \ 
            & 3.37  &   2.30    &  0.47   &   \  \ 
            & 0.91  &   0.24    &  0.31    \\
\hline \hline 
\end{tabular}
\end{center}
\label{CV-g}
\end{table*}
\begin{table*}
\caption[t]{Our calculated ${\mathbf U}^{{\rm cRPA}}$, ${\mathbf U}^{{\rm D\!M\!F\!T}}$, and ${\mathbf W}$ at fillings $n$=1.0-5.0 ({\it Tokyo Ab initio Program Package}). These data are used in Fig.~2 in the main text. The ${\mathbf U}^{{\rm no\mathchar`-wing}}$ data are also listed. For the definition of ${\mathbf U}^{{\rm no\mathchar`-wing}}$, see the main text.}
\begin{center}
\begin{tabular}{r ccc c ccc c ccc c ccc}
\hline \hline \\ [-5pt]
 & \multicolumn{3}{c}{${\mathbf U}^{{\rm cRPA}}$} & 
 & \multicolumn{3}{c}{${\mathbf U}^{{\rm D\!M\!F\!T}}$} &  
 & \multicolumn{3}{c}{${\mathbf U}^{{\rm no\mathchar`-wing}}$} & 
 & \multicolumn{3}{c}{${\mathbf W}$} \\ 
 \hline
 & $U$ & $U'$ & $J$ & 
 & $U$ & $U'$ & $J$ &  
 & $U$ & $U'$ & $J$ & 
 & $U$ & $U'$ & $J$ \\ 
 \hline 
 $n$=1.0  \ \  
            & 3.39  &   2.34    &  0.47   & \ \ 
            & 3.33  &   2.27    &  0.47   & \ \    
            & 3.30  &   2.24    &  0.47   & \ \ 
            & 0.97  &   0.25    &  0.33   \\
 $n$=1.5  \ \  
            &  3.47 &   2.41    &  0.47   & \ \ 
            &  4.01 &   2.93    &  0.48   & \ \   
            &  3.36 &   2.29    &  0.48   & \ \ 
            &  0.80 &   0.16    &  0.29   \\
 $n$=2.0  \ \  
            &  3.65  &  2.59     &  0.46   & \ \ 
            &  4.74 &   3.63    &  0.47   & \ \      
            &  3.41 &   2.34    &  0.47   & \ \  
            &  0.68 &   0.11    &  0.26   \\ 
 $n$=2.5  \ \  
            &  3.72 &   2.65    &  0.46   & \ \ 
            &  4.58 &   3.48    &  0.47   & \ \   
            &  3.23 &   2.16    &  0.47   & \ \ 
            &  0.59 &   0.07    &  0.24   \\ 
 $n$=3.0  \ \  
            &  3.83 &   2.75    &  0.45   & \ \ 
            &  4.33 &   3.23    &  0.46   & \ \      
            &  3.14 &   2.07    &  0.46   & \ \ 
            &  0.53 &   0.06    &  0.22   \\  
 $n$=3.5  \ \  
            &  3.89 &   2.81    &  0.45   & \ \ 
            &  3.85 &   2.76    &  0.45   & \ \      
            &  3.01 &   1.96    &  0.45   & \ \ 
            &  0.49 &   0.04    &  0.20   \\ 
 $n$=4.0  \ \  
            &  3.93 &   2.85    &  0.44   & \ \ 
            &  3.39 &   2.32    &  0.44   & \ \      
            &  3.02 &   1.96    &  0.44   & \ \ 
            &  0.47 &   0.04    &  0.20   \\  
 $n$=4.5  \ \  
            &  3.98 &   2.90    &  0.44   & \ \ 
            &  3.05 &   2.00    &  0.43   & \ \      
            &  2.94 &   1.90    &  0.43   & \ \ 
            &  0.50 &   0.05    &  0.20   \\   
 $n$=5.0  \ \  
            &  4.06 &   2.97    &  0.43   & \ \ 
            &  3.58 &   2.50    &  0.43   & \ \      
            &  2.75 &   1.71    &  0.42   & \ \ 
            &  0.62 &   0.08    &  0.24   \\ 
\hline \hline 
\end{tabular}
\end{center}
\label{U-n}
\end{table*}

%% file: main.bbl
\begin{thebibliography}{999}
\bibitem{note2}The quantity $\tilde{U}+\frac{\tilde{U}}{1+\tilde{U}A}$ is nearly equal to the fully screened Coulomb interaction, which is a smooth function of filling.
\end{thebibliography}
\begin{thebibliography}{99}
\bibitem[S5]{note}
When we consider the impurity problem in the real space, the translational
symmetry of the primitive lattice is broken. So, we have to express all the
quantities in terms of the periodicity of the superlattice. Note that the
system has still a translational symmetry of the superlattice, due to the
periodic boundary condition. The present derivation follows the superlattice
formulation. On the other hand, the fully screened RPA Coulomb interaction
${\mathbf W}$ needed in the ${\mathbf U}^{{\rm D\!M\!F\!T}}$ calculation can
be calculated in the original periodicity of the primitive lattice, because
${\mathbf W}$ itself is free from the impurity problem. In the present work,
we calculate ${\mathbf W}$ for the original primitive lattice.
\bibitem[S9]{noteS2}

We first evaluated the {\bf k} integral by the generalized tetrahedron
method \cite{tetra} in a finer mesh with the interpolation technique, then we
performed the sum over {\bf q} in the original mesh.
\end{thebibliography}
